\documentstyle[12pt,amsfonts]{article}
\parskip=\medskipamount

\def\appendix#1{
  \addtocounter{section}{1}
  \setcounter{equation}{0}
  \renewcommand{\thesection}{\Alph{section}}
 \section*{Appendix \thesection\protect\indent
 \parbox[t]{11.715cm} {#1}}
 \addcontentsline{toc}{section}{Appendix \thesection\ \ \ #1}
  }
\renewcommand{\thefootnote}{\fnsymbol{footnote}}

\newcommand {\cA}{{\cal A}}

\newcommand {\cD}{{\cal D}}

\newcommand {\cF}{{\cal F}}

\newcommand {\cN}{{\cal N}}

\newcommand {\cY}{{\cal Y}}

\def\a{\alpha}
\def \bi{\bibitem}

\def\b{\beta}
\def\c{\chi}
\def\d{\delta}
\def\e{\epsilon}
\def\f{\phi}
\def\g{\gamma}
\def\G{\Gamma}

\def\l{\lambda}
\def\m{\mu}
\def\n{\nu}

\def\p{\pi}

\def\r{\rho}
\def\s{\sigma}
\def\t{\tau}

\def\x{\xi}
\def\z{\zeta}
\def\D{\Delta}
\def\F{\Phi}
\def\J{\Psi}
\def\L{\Lambda}
\def\O{\Omega}

\newcommand{\pa}{\partial}                           
\newcommand{\hf}{\frac12}

\newcommand{\vf}{\varphi}
\newcommand{\sect}[1]{\setcounter{equation}{0}\section{#1}}

\newcommand{\be}{\begin{equation}}
\newcommand{\ee}{\end{equation}}
\newcommand{\bea}{\begin{eqnarray}}
\newcommand{\eea}{\end{eqnarray}}
\newcommand{\non}{\nonumber}
\begin{document}

\begin{titlepage}
\thispagestyle{empty}

\begin{flushright}
hep-th/0203236 \\
March, 2002
\end{flushright}
\vspace{5mm}

\begin{center}
{\Large \bf Quantum metamorphosis  of conformal symmetry\\
in N = 4 super Yang-Mills theory}
\end{center}
\vspace{3mm}

\begin{center}
{\large
S.M. Kuzenko and I.N. McArthur
}\\
\vspace{2mm}

${}$\footnotesize{
{\it
Department of Physics, The University of Western Australia\\
Crawley, W.A. 6009. Australia}
} \\
{\tt  kuzenko@cyllene.uwa.edu.au},~
{\tt mcarthur@physics.uwa.edu.au}
\vspace{2mm}

\end{center}
\vspace{5mm}

\begin{abstract}
\baselineskip=14pt
In gauge theories, not all rigid symmetries of the classical
action can be maintained manifestly in the quantization procedure,
even in the absence of anomalies. If this occurs for an anomaly-free
symmetry, the effective action is invariant under a transformation
that differs from its classical counterpart by quantum corrections.
As shown by Fradkin and Palchik years ago, such a phenomenon
occurs for conformal symmetry in quantum Yang-Mills theories with
vanishing beta function, such as the $\cN=4$ super Yang-Mills theory.
More recently, Jevicki et al demonstrated that the quantum
metamorphosis of conformal symmetry sheds light on the nature
of the AdS/CFT correspondence. In this paper, we derive the conformal
Ward identity for the bosonic sector of the $\cN=4$ super Yang-Mills
theory using the background field method. We then compute the leading
quantum modification of the conformal transformation for a specific Abelian
background which is of interest in the context of the
AdS/CFT correspondence. In the case of scalar fields, our final
result agrees with that of  Jevicki et al. The resulting vector
and scalar transformations coincide with those which are
characteristic of a D3-brane embedded in $AdS_5 \times S^5$.
\end{abstract}

\vfill
\end{titlepage}

\newpage
\setcounter{page}{1}

\renewcommand{\thefootnote}{\arabic{footnote}}
\setcounter{footnote}{0}
\sect{Introduction}

$\cN=4$ super Yang-Mills theory was  popular in the
1980's as the unique maximally supersymmetric Yang-Mills
theory in four space-time dimensions, and
the first ultraviolet-finite quantum field theory ever constructed.
More recently, it has become the subject of immense
scrutiny in the context of
the AdS-CFT correspondence \cite{M,GKP,Witten,AGMOO}.
Yet there is another interesting field theoretic
aspect of this dynamical system
which  has so far not received much attention,  except in \cite{JKY},
and which sheds light on the origin of
the AdS-CFT correspondence.
Quantum conformal invariance in the
$\cN=4$ super Yang-Mills theory turns out to be
a nontrivial deformation\footnote{By ``quantum deformation''
of a rigid symmetry we understand modifications
to the field transformations due to quantum corrections
in such a way that the rigid symmetry algebra remains intact.
This should not be confused with the term ``quantum deformation''
in the context of quantum groups.}
of the linear conformal
symmetry of the classical action, and this fact has
interesting implications.

In 1984, it was shown by Fradkin and Palchik  \cite{FP1}
(see also \cite{Pal,FP2}) that the generating functional
in conformally invariant quantum non-Abelian gauge theories
is not invariant under standard (linearly realized)
special conformal transformations; such an invariance
is only consistent  with a purely longitudinal
two-point function, $\langle A_m (x_1) \,A_n (x_2) \rangle
\propto \pa_m \pa_n \ln \,(x_1 - x_2)^2$.
These theories are, however, invariant under
{\it deformed} special conformal transformations consisting of
a combination of  conformal transformations
and  compensating field-dependent gauge transformations;
the conformal Ward identity associated with the deformed symmetry
leads to the correct transverse propagator.

In 1998, Jevicki {\it et al} \cite{JKY} applied and extended
the Fradkin-Palchik construction to address a question
that may be formulated as follows:
Where are the branes implied by the AdS/CFT duality
within the framework  of super Yang-Mills theory?
Using a derivative expansion
of the effective action, they computed the leading quantum
deformation of the conformal transformation law
of the Higgs fields $Y_\m$, with $\m=1,\ldots,6$,
which trigger the spontaneous breakdown of
the gauge group $S(N+1)$ to $SU(N)\times U(1)$
in the $\cN=4$ super Yang-Mills theory.
The deformed transformation is
\be
\d Y_\m = -\d x^m \, \pa_m Y_\m +2(b \cdot x )\, Y_\m ~,
\qquad \d x^m = b^m x^2 -2x^m \,(b \cdot x) \,
+ b^m \, \frac{R^4}{Y^2}~,
\label{Higgs}
\ee
where $Y^2 = Y_\m Y_\m$ and  $R^4 = N\, g_{\rm YM}^2 / (2 \p^2)$,
and  coincides with that of the transverse degrees of freedom
of a D3-brane embedded in $AdS_5 \times S^5$. In conjunction
with the requirement of $SO(6)$ invariance
and some non-renormalization theorems in $\cN=4$ SYM, the above
transformation uniquely fixes the part of
the low energy effective action for the D3-brane
which depends on the scalar fields and their first derivatives only
\cite{M},
\be
S = - \frac{1}{ g_{\rm YM}^2 R^4} \int {\rm d}^4x \,
{Y^4}\,
\left( \sqrt{ - {\rm det} ( \eta_{m n}
+  R^4 \, \partial_{m}Y \cdot \partial_{n} Y \,/ Y^4 )}
- 1\right) ~.
\label{Higgs-action}
\ee

The coupling constant $g_{\rm YM}^2$ can be treated
as a loop-counting parameter in $\cN=4$
super Yang-Mills theory. In this respect,
the $R^4$ dependent term in the transformation $\d Y$ in eq. (\ref{Higgs})
is just a one-loop quantum deformation.
On the other hand, the action (\ref{Higgs-action})
is the result of summing up  quantum corrections to all loop orders.
Therefore, even the one-loop symmetry deformation
contains essential information about
the structure of the effective action
at higher loops! This is one of the main reasons  why we consider
it important to pursue the study
of quantum deformations of (super) conformal
symmetry in finite $\cN=2, \,4$ super Yang-Mills theories.

The story with conformal symmetry is an example of the more general
phenomenon of quantum deformations of rigid symmetries in gauge
theories, which in fact embraces two different aspects: (i)
deformations before gauge fixing; and (ii) deformations after
gauge fixing. Point (i)  concerns the problem of extending
any rigid symmetry of a classical gauge invariant action
to the ghosts and the antifields, in the framework of the
Batalin-Vilkovisky quantization scheme \cite{BV}
and its extensions, so as to leave
the solution of the master equation invariant.
It is always possible to construct a solution
to this problem \cite{BHW} in which the global symmetry
is (antibracket) canonically realized.
Point (ii) has been analysed in a rather general setting
by van Holten \cite{vH}, and here we  simply quote his
formulation of the problem: it does not hold that ``any rigid symmetry
of a classical action can always be maintained manifestly
in the Faddeev-Popov--BRST quantization scheme, even in
the absence of anomalies.'' Examples include
on-shell supersymmetry \cite{dWF},
conformal symmetry \cite{FP1} and its supersymmetric extensions.
According to the analysis of \cite{vH}, the problem of keeping
the rigid symmetries manifest at the quantum level is essentially
equivalent to finding covariant gauge conditions.
In the case of conformal symmetry, such gauge conditions
do not exist \cite{FP1} and any special conformal
transformation has to be accompanied by a field-dependent
nonlocal gauge transformation  in order to restore the gauge
slice.

The present paper is organized as follows.
In section 2, we give a general discussion of
quantum deformations of global symmetries
(for a  general gauge theory)
in the framework of  DeWitt--Faddeev-Popov--BRST
quantization. This section is of  a review nature
and  consists largely  of variations on themes
suggested by Fradkin and Palchik, van Holten and others.
In section 3, we derive the conformal Ward identity
in the $\cN=4$ super Yang-Mills theory within
the background field method. The analysis in this paper is
restricted to the (full) bosonic sector of the theory,
simply because the fermions are not relevant
for the subsequent considerations; they can be included by simple
extension. The content of section 3
significantly extends the  earlier results for scalar fields
given in \cite{JKY}. In addition, our derivation of the conformal
Ward identity is more rigorous. It is worth mentioning
that the content of section 3 can  easily be
generalized to the case of finite $\cN=2$ super Yang-Mills
models. In section 4, we compute the leading quantum deformation
of the conformal transformation for a specific Abelian
background which is of interest in the context of the
AdS/CFT correspondence. The resulting vector and scalar
transformations coincide with those
characteristic of a D3-brane embedded
in $AdS_5 \times S^5$.  The results  and interesting
open problems are then discussed in section 5.

\sect{Symmetries of the effective action}

In this section, we use DeWitt's condensed notation
\cite{DeWitt67} (see also \cite{DeWitt2}),
and for simplicity restrict attention
to the case of bosonic gauge theories.
Let $S[\F]$ be the action of an irreducible gauge theory
(following the terminology of \cite{BV})
involving bosonic fields $\F^i$.
By definition, the gauge generators $R^i{}_\a [\F]$
give rise to  Noether identities
\be
S_{,i}[\F] \, R^i{}_\a [\F] \equiv 0~,
\ee
and in what follows they are assumed to form a closed algebra,
\be
R^i{}_{\a,j}[\F]\, R^j{}_\b [\F]
- R^i{}_{\b,j}[\F]\, R^j{}_\a [\F]
= R^i{}_\g [\F]\,f^\g{}_{\a \b} [\F]~,
\ee
together with additional requirements
\be
R^i{}_{\a,i} [\F] =0~, \qquad f^\b{}_{\a \b} [\F] = 0~.
\label{ad-req}
\ee
It will be also assumed that the gauge transformations
\be
\d \F^i = R^i{}_\a [\F] \, \d\z^\a~,
\ee
with $\d \z^\a$ arbitrary parameters of compact support,
span the gauge freedom of the theory -- that is, if $\F_0 $
is a stationary point  of the action,
$S_{,i}[\F_0] =0$, then the equality
$R^i{}_\a [\F_0] \d \z^\a = 0$ implies
$\d \z^\a = 0$.

Under the above assumptions, the in-out vacuum amplitude
is known to have a  functional integral representation
of the form
\be
\langle {\rm out} | ~{\rm in} \rangle =
N \int {\rm d} \F \, {\rm Det}  (F[\F]) \,
{\rm e}^{ {\rm i} ( S[\F] + S_{\rm GF} [\c [\F ]] )}~,
\label{in-out}
\ee
where $\c^\a [\F]$ are gauge conditions such that the operator
\be
F^\a{}_\b [\F] \equiv \c^\a{}_{, i}[\F] R^i{}_\b [\F]
\ee is non-singular at
$\F_0$. The gauge fixing functional $ S_{\rm GF} [\c ]$
is chosen in such a way
that the action $S[\F] + S_{\rm GF} [\c [\F ]] $ is no longer gauge
invariant. In perturbation theory,
it is customary to choose $ S_{\rm GF} [\c ]$
to be of Gaussian form,
$ S_{\rm GF} [\c ] = \hf \c^\a \eta_{\a \b} \c^\b$,
with $\eta_{\a \b}$ a constant non-singular symmetric matrix.

The  in-out vacuum amplitude (\ref{in-out}) is independent
of the  choice of $\c$,
$ \langle {\rm out} | ~{\rm in} \rangle_{\c + \d \c}
= \langle {\rm out} | ~{\rm in} \rangle_\c$,
with $\d \c^\a [\F] $ a small deformation of the gauge conditions.
It is important for  subsequent considerations
to recall an old proof\footnote{Similar arguments can be used
to establish the independence of the vacuum to vacuum amplitude on
the functional form of $ S_{\rm GF}$; see \cite{DeWitt1}.
The inclusion of Nielsen-Kallosh ghosts \cite{NK}
may be important.} of this fact
due to DeWitt \cite{DeWitt67,DeWitt1}.
In the functional integral
\be
\langle {\rm out} | ~{\rm in} \rangle_{\c + \d \c}
=
N \int {\rm d} \tilde{\F} \,
{\rm Det}  (F[\tilde{\F}] +\d F[ \tilde{\F}]) \,
{\rm e}^{ {\rm i} ( S[\tilde{\F}] +
S_{\rm GF} [\c [\tilde{\F} ] + \d \c [\tilde{\F}]] )}~,
\ee
with $\d F^\a{}_\b [\F] = \d \c^\a{}_{,i}[\F] R^i{}_\b [\F]$,
we make a replacement of integration variables
\be
\tilde{\F}^i = \F^i - R^i{}_\a [\F] \d \z^\a [\F], \qquad
\d \z^\a [\F] = (F^{-1} [\F])^\a{}_\b  \, \d \c^\b [\F]
\label{field-dep}
\ee
chosen so that
$$
S[\tilde{\F}] + S_{\rm GF} [\c [\tilde{\F} ] + \d \c [\tilde{\F}]]
=  S[\F] + S_{\rm GF} [\c [\F ]].
$$
On the other hand, direct calculations yield
\bea
{\rm d} \tilde{\F} \,
{\rm Det}  (F[\tilde{\F}] +\d F[ \tilde{\F}])
&=& {\rm d}\F \, {\rm Det}  (F[\F] ) \, \Big\{
1 - R^i{}_{\a,i} [\F] \d \z^\a[\F]
-f^\b{}_{\b \a} [\F] \d \z^\a [\F]  \Big\} \non \\
&=& {\rm d}\F \, {\rm Det}  (F[\F] )~,
\eea
as a consequence of (\ref{ad-req}).
Similar considerations can be used to show that
the correlation function
$\langle {\rm out} | \J[\F] |~{\rm in} \rangle$ of a gauge invariant
functional $\J[\F]$, with $\J_{,i} [\F] \, R^i{}_\a [\F]= 0$, is
not dependent on the gauge choice.

Next, we turn our attention to the effective action of the theory,
\be
\G[\f] = \big( W[J] -  J_i \, \f^i \big)\big|_{J= J[\f]}~, \qquad
\f^i = \frac{ \d}{ \d J_i} W[J]~,
\ee
with $W[J]$ the generating functional of connected Green's
functions,
\be
{\rm e}^{{\rm i} W[J]} =
N \int {\rm d} \F \, {\rm Det}  (F[\F]) \,
{\rm e}^{ {\rm i} ( S[\F] + S_{\rm GF} [\c [\F ]]
+ J_i \F^i)}~.
\label{gen-fun}
\ee
The effective action   depends explicitly on the choice of
the gauge condition $\c[\F]$,
unlike the $S$-matrix following from $\G[\f]$.
It is not, however, the issue of gauge dependence which
is the point of  concern here.
Suppose the classical action is invariant,
$S[ \F + \e \, \O[\F] ] = S[\F]$,
under a rigid transformation
\be
\d \F^i = \e \, \O^i[\F] ~,
\label{glob}
\ee
with $\e$ an infinitesimal constant parameter.
We will analyse the implications of this classical symmetry
for the effective action; see Ref. \cite{vH}
for a similar earlier treatment.
In what follows, some additional properties of the
structure of the gauge and global transformations
will be assumed, namely
\bea
\O^i{}_{,i} [\F] & = & 0 ~, \label{div} \\
R^i{}_{\a,j} [\F] \,\O^j [\F] -
\O^i{}_{,j}[\F] \, R^j{}_\a [\F]
&=& R^i{}_{\b} [\F] \,f^\b{}_\a [\F] ~, \label{algebra} \\
f^\a{}_\a [\F] &=& 0~. \label{contr}
\eea
Eq. (\ref{div}) ensures that the Jacobian of the transformation
$\F^i \to \F^i +\e \, \O^i[\F]$ is equal to one.
Eq. (\ref{algebra}) implies that the commutator of a gauge
transformation with a global symmetry transformation is a gauge
transformation.

To understand the manifestations of the symmetry
(\ref{glob}) at the quantum level,
we make the  change of variables
\be
\F^i = \tilde{\F}^i + \e \, \O^i[\tilde{\F}]
\ee
in the right hand side of (\ref{gen-fun}).
Using eqs. (\ref{algebra}) and (\ref{contr}),
one then obtains
\bea
F^\a{}_\b [\F] &=& F^\a{}_\b [ \tilde{\F} ]
+ \d_\e F^\a{}_\b [\tilde{\F}]
+  \e\, F^\a{}_\g [ \tilde{\F} ] \,
f^\g{}_\b [\tilde{\F}] ~, \\
{\rm Det} ( F[\F])
&=& {\rm Det} ( F[\tilde{\F}]  + \d_\e F[\tilde{\F}]) ~,
\eea
where
$$
\d_\e F^\a{}_\b [\tilde{\F}] =
\d_\e \c^\a{}_{,i}[\tilde{\F}] \, R^i{}_\b [\tilde{\F}] ~, \qquad
\d_\e \c^\a [\tilde{\F}] =
\e \, \c^\a{}_{,i} [ \tilde{\F}] \, \O^i[ \tilde{\F} ] ~.
$$
As a result, eq. (\ref{gen-fun}) becomes
\bea
{\rm e}^{{\rm i} W[J]} &=&
N \int {\rm d} \tilde{\F} \,
{\rm Det}  ( F[\tilde{\F}] + \d_\e F[\tilde{\F}] ) \non \\
& \times &
\exp \, {\rm i} \Big\{ S[\tilde{\F}] +
S_{\rm GF} [\c [\tilde{\F }] +\d_\e \c [\tilde{\F}] ]
+ J_i (\tilde{\F}^i + \e\, \O^i[\tilde{\F}])  \Big\}~.
\eea
In the functional integral obtained,
we can then change variables according to
the rule  (\ref{field-dep})
with $\d \c [\F] = \d_\e \c [\F] $.
This leads to the  Ward identity
\be
\G_{,i} [\f] \; \langle \O^i [\F] \rangle
~=~  \G_{,i} [\f] \; \langle R^i{}_\a [\F] \,
(F^{-1} [\F] )^\a{}_\b \,
\c^\b{}_{,j} [\F] \, \O^j [\F ] \rangle ~,
\label{ST1}
\ee
where we  have used the fact that $J_i = -\G_{,i}[\f]$.
In the case of non-gauge theories, the right hand side
of eq. (\ref{ST1}) vanishes; see, for example, Weinberg's book
\cite{Wei}. In eq. (\ref{ST1}), the symbol
$\langle ~~~ \rangle $ denotes
the quantum average in the presence of the source $J =  J[\f]$,
\be
\langle A[ \F ] \rangle
~=~ {\rm e}^{-{\rm i} W[J]} \,
N \int {\rm d} \F \, A[\F] \,{\rm Det}  (F[\F]) \,
{\rm e}^{ {\rm i} ( S[\F] + S_{\rm GF} [\c [\F ]]
+ J_i \F^i)}~.
\ee

${}$For a large class of gauge theories, the above Ward
identity can be brought into a simpler form.
Suppose that the gauge fixing functional is invariant,
$S_{\rm GF} [\c + \d_\L \c] = S_{\rm GF} [\c ]$,
under a linear homogeneous transformation
\be
\d_\L \c^\a = \e \, \L^\a{}_\b \c^\b~,
\label{lambda}
\ee
with $\L^\a{}_\b$ a field independent operator. Since
$\L$ is field independent, we have
\bea
{\rm e}^{{\rm i} W[J]} =
N \int {\rm d} \tilde{\F} \,
{\rm Det}  ( F[\tilde{\F}] + \d_\L F[\tilde{\F}] ) \,
{\rm e}^{ {\rm i} ( S[\tilde{\F}] +
S_{\rm GF} [\c [\tilde{\F }] +\d_\L \c [\tilde{\F}] ]
+ J_i \tilde{\F}^i  )} ~, \non
\eea
where $\d_\L F^\a{}_\b [\tilde{\F}] =
\d_\L \c^\a{}_{,i}[\tilde{\F}] \, R^i{}_\b [\tilde{\F}]$.
Replacing  the integration variables according to
the rule  (\ref{field-dep}) with $\d \c [\F] = \d_\L \c [\F] $,
one then obtains
\be
\G_{,i} [\f] \; \langle R^i{}_\a [\F] \,
(F^{-1} [\F] )^\a{}_\b \, \L^\b{}_\g  \,
\c^\g [\F]  \rangle ~=~0~.
\label{something}
\ee

Our last assumption concerns the behaviour of
the gauge conditions under the symmetry transformation. We assume
\be
\d_\e \c^\a [\F] \equiv
\e \, \c^\a{}_{,i} [\F] \, \O^i[\F]
=\e \,\Big( \L^\a{}_\b \, \c^\b [\F] + \r^\a [\F] \Big)~, \qquad
\r^\a [\F] \neq 0~,
\label{gauge-cond-tran}
\ee
where the homogeneous term on the right hand side
leaves $S_{\rm GF}[\c] $ invariant,
$ S_{\rm GF}[\c^\a + \e \, \L^\a{}_\b \, \c^\b]
= S_{\rm GF}[\c^\a ]$. In this case, the Ward
identity (\ref{ST1}) is equivalent to
\be
\G_{,i} [\f] \; \langle \O^i [\F] \rangle
~=~  \G_{,i} [\f] \; \langle R^i{}_\a [\F] \,
(F^{-1} [\F] )^\a{}_\b \,
\r^\b [\F ] \rangle ~.
\label{ST2}
\ee

It is worth discussing the transformation law
(\ref{gauge-cond-tran}). The gauge conditions
$\c^\a [\F] = 0$ break the gauge invariance
and single out a unique representative from each
gauge orbit. For most global symmetries, there exist
{\it covariant}  gauge conditions - that is, they
can be chosen in such a way
that  $\c^\a [\F] $ transforms as in eq.  (\ref{gauge-cond-tran})
but with $\r^\a [\F] =0$. In this case, the global symmetry
leaves the gauge slice $\c^\a [\F] = 0$  invariant.
However, for some symmetries, such as conformal invariance,
there is no way to eliminate the inhomogeneous term in
(\ref{gauge-cond-tran}), and the symmetry transformation
does not leave the gauge conditions $\c^\a [\F] = 0$
invariant. In such a situation,
a non-trivial symmetry deformation occurs at the quantum
level, as follows from eq. (\ref{ST2}).
Indeed,  consider the simplest situation of
a linearly realized classical symmetry,
$\O^i [\F] = \O^i{}_j \, \F^j$. In this case, the left hand side in
(\ref{ST2}) is simply $\G_{,i} [\f] \, \O^i [\f]$, and hence
eq. (\ref{ST2}) can be interpreted as the invariace
condition under deformed transformations of the form
$\d \f^i ~= ~ \e \,\O^i [\f] ~+~ {\rm quantum ~corrections}.$

It is instructive to re-derive the above results
in the BRST approach \cite{BRST}, in which
eq. (\ref{gen-fun}) is replaced by
\be
{\rm e}^{{\rm i} W[J]} =
N \int {\rm d} \F \, {\rm d} {\bar C} \,{\rm d} C \,
{\rm e}^{ {\rm i} ( S_{\rm eff}[\F, {\bar C},C; \c[\F] ]
+ J_i \F^i)}~,
\label{gen-fun2}
\ee
where
\be
S_{\rm eff}[\F, {\bar C},C; \c[\F]] = S[\F]
+ S_{\rm GF} [\c [\F ]]
+{\bar C}_\a  \c^\a{}_{,i} [\F] R^i{}_\b [\F] C^\b~,
\ee
with $\bar C_\a$ and $C^\a$ the Faddeev-Popov ghosts.
The action $S_{\rm eff}$ is invariant under the following
BRST transformation
\be
\d \F^i = R^i{}_\a [\F] C^\a \, \l~, \quad
\d C^\a = \hf f^\a{}_{\b \g } [\F] C^\g C^\b \, \l ~,
\quad
\d {\bar C}_\a =  S_{{\rm GF} , \a} [\c[\F]] \, \l~,
\label{BRST}
\ee
 with $\l$ a constant anticommuting parameter.
This BRST transformation leaves  the integration
measure ${\rm d} \F  {\rm d} {\bar C} {\rm d} C$
in (\ref{gen-fun2}) invariant as a consequence of (\ref{ad-req}).

Inspired by  \cite{BFV},  we make a BRST-like change  of variables
with a field dependent parameter in the right hand side of
(\ref{gen-fun2}),
\bea
\F^i  \to  \F^i  + R^i{}_\a [\F] C^\a \, \l~, \quad
C^\a \to C^\a &+& \hf f^\a{}_{\b \g } [\F] C^\g C^\b \, \l ~,
\quad
{\bar C}_\a \to {\bar C}_\a +  S_{{\rm GF} , \a} [\c[\F]] \, \l~,
\non \\
\l &=& -{\bar C}_\a \, \d \c^\a [\F] ~,
\label{field-dep-BRST}
\eea
where $\d \c^\a [\F]$ are arbitrary variations
of the gauge conditions. This transformation obviously leaves
$S_{\rm eff}$ invariant, but the corresponding Jacobian is
now non-trivial, and eq. (\ref{gen-fun2}) turns into
\bea
{\rm e}^{{\rm i} W[J]} &=&
N \int {\rm d} \F \, {\rm d} {\bar C} \,{\rm d} C \,
{\rm e}^{ {\rm i} ( S_{\rm eff}[\F, {\bar C},C; \c[\F]
+\d \c [\F]]
+ J_i \F^i)} \non \\
& \times & \Big( 1 -{\rm i}\, J_i \, R^i{}_\a [\F] \, C^\a \,
{\bar C}_\b \, \d \c^\b [\F] \Big)~.
\eea
${}$For $J= 0,$ this relation is equivalent to the gauge-independence
of the in-out vacuum amplitude.

Given a rigid symmetry defined by eqs. (\ref{glob})
-- (\ref{contr}), one can consider the following change of variables
\be
\F^i  \to  \F^i  + \e \,\O^i[\F] ~, \quad
C^\a \to C^\a - \e \,f^\a{}_{\b  } [\F] \,C^\b  ~,
\quad
{\bar C}_\a \to {\bar C}_\a
\ee
in the right hand side of (\ref{gen-fun2}).
This leads to
\bea
{\rm e}^{{\rm i} W[J]} &=&
N \int {\rm d} \F \, {\rm d} {\bar C} \,{\rm d} C \,
{\rm e}^{ {\rm i} ( S_{\rm eff}[\F, {\bar C},C; \c[\F]
+\d_\e \c [\F]]
+ J_i \F^i)} \non \\
& \times & \Big( 1 +{\rm i}\, J_i \, \O^i [\F] \Big)~,
\eea
where $\d_\e \c^\a [\F] =
\e \, \c^\a{}_{,i} [\F] \, \O^i[\F]$.
The change of the gauge conditions in $S_{\rm eff}$
can be compensated by  a BRST-like
transformation (\ref{field-dep-BRST})
with $\l ={\bar C}_\a \, \d_\e \c^\a [\F]$.
As a result, one obtains a new realization
of the Ward identity (\ref{ST1}),
\be
\G_{,i} [\f] \; \langle \O^i [\F] \rangle
~=~  -\G_{,i} [\f] \; \langle R^i{}_\a [\F] \,
C^\a\, {\bar C}_\b \,
\c^\b{}_{,j} [\F] \, \O^j [\F ] \rangle ~.
\label{ST3}
\ee
Similarly, the BRST counterparts of eqs.
(\ref{something}) and (\ref{ST2}) are obtained
via the substitution
$(F^{-1} [\F])^\a{}_\b \to - C^\a \,{\bar C}_\b$.
In the BRST approach, it is worth pointing out that
the transformation (\ref{lambda}), which leaves
$S_{\rm GF} [\c]$ invariant,
 becomes
\be
\d_\L \c^\a = \e \, \L^\a{}_\b \,\c^\b~, \quad
\d_\L C^\a  = 0~, \quad
\d_\L {\bar C}_\a = - \e \, {\bar C}_\b \,
\L^\b{}_\a ~,
\ee
which does not change the  functional
$ S_{\rm GF} [\c [\F ]]
+{\bar C}_\a  \c^\a{}_{,i} [\F] R^i{}_\b [\F] C^\b$.

\sect{Conformal Ward identity in \mbox{$\cN=4$} SYM}

As is well known, the $\cN=4$ super Yang-Mills theory can be
obtained by plain dimensional reduction
from  super Yang-Mills theory in ten dimensions:
\be
S = -\frac{1}{4g^2}\int {\rm d}^{10} x \; {\rm tr} \Big(
F^{MN} F_{MN} - 2 {\rm i} \,\bar{\J} \G^M D_M \J \Big)~,
\label{10Daction}
\ee
where $F_{MN} = \pa_M A_N - \pa_N A_M +{\rm i} [A_M , A_N]$.
In the present paper, we are interested in the bosonic sector
of the $\cN=4$ super Yang-Mills theory described by
fields $A_M = (A_m, Y_\m) $, where $m=0,1,2,3$ and $\m=1,\ldots, 6$.
The classical action reads
\be
S [A,Y] = -\frac{1}{4g^2}\int {\rm d}^{4} x \; {\rm tr} \Big(
F^{mn} F_{mn} + 2 D^m Y_\m D_m Y_\m
- [Y_\m, Y_\n]\, [Y_\m, Y_\n] \Big)~,
\label{action}
\ee
with $D_m = \pa_m +{\rm i} A_m$, and is invariant under standard
gauge transformations
\be
\d A_m = - D_m \t = - \pa_m \t -{\rm i}\,[A_m, \t]~,
\qquad \d Y_\m = {\rm i}\, [\t, Y_\m]~.
\ee

The action (\ref{action}) is also invariant under arbitrary conformal
transformations
\be
-\d_{\rm c} A_m = \x A_m +\hat{K}_m{}^n A_n + \s A_m~, \qquad
-\d_{\rm c} Y_\m =  \x Y_\m +\s Y_\m~,
\label{conf}
\ee
where $\x= \x^m \pa_m$ is a conformal Killing vector field,
\be
\pa_m \x_n + \pa_n \x_m = 2\eta_{mn}\, \s~, \qquad
\s \equiv \frac{1}{4}\pa_m \x^m~, \qquad
\hat{K}_{mn} \equiv \hf (\pa_m \x_n -  \pa_n \x_m)~.
\ee
The general solution to the conformal
Killing equation is
\be
\x^m = a^m + \l x^m +K^m{}_n x^n +b^m x^2 - 2x^m (b \cdot x)~,
\qquad K_{mn}= -  K_{nm}~,
\ee
where
\be
\s = \l - 2 (b \cdot x)~.
\ee
Our goal below will be to analyse how these conformal transformations
(\ref{conf}) are deformed at the quantum level.

We will quantize the $\cN=4$ SYM theory in the framework
of the background field method
(see \cite{DeWitt67,tH,DeWitt3,Ab,Boul,Hart} and references therein),
by splitting the dynamical variables
$\F^i = (A_m, Y_\m)$ into the sum of {\it background}
fields $\f^i = (\cA_m , \cY_\m)$ and {\it quantum}
fields $\vf^i = (a_m, y_\m)$.  The classical action
$S[\f + \vf] = S[\cA +a, \cY +y]$ is then invariant under
{\it background} gauge transformations
\bea
\d \cA_m = - \cD_m \t ~, & \qquad  &
\d Y_\m = {\rm i} [\t , \cY_\m ] ~, \non \\
\d a_m = {\rm i} [\t , a_m ]~, & \qquad &
\d y_\m = {\rm i} [\t , y_\m ]~;
\eea
and {\it quantum} gauge transformations
\bea
\d \cA_m = 0~, & \qquad & \d \cY_\m =  0 ~, \non \\
\d a_m = - \cD_m - {\rm i} [a_m , \t ] ~, & \qquad &
\d y_\m = {\rm i} [\t , \cY_\m  +y_\m ] ~,
\label{quantum}
\eea
with $\cD_m $ the background covariant derivatives.
The background field quantization procedure
consists of fixing the quantum gauge freedom
while  keeping  the background gauge invariance intact
by means of  background  covariant gauge conditions.
The effective action is given by the sum of all 1PI Feynman graphs
 which are vacuum with respect to the quantum fields.

In 't Hooft gauge, the gauge conditions $\c^\a$ are
\be
\c  = \cD^m a_m + {\rm i} \,[\cY_\m , y_\m ] ~,
\label{g-c}
\ee
and the gauge fixing functional, $S_{\rm GF}$, is
\be
S_{\rm GF} [\c] = -\frac{1}{ 2g^2} \int {\rm d}^4 x \;
{\rm tr} \, \c^2 ~.
\ee
Under the quantum gauge transformations (\ref{quantum}),
\be
\d_{\rm quantum} \c =
-\cD^m ( \cD_m \t +{\rm i}\, [a_m, \t] )
+ [ \cY_\m , [\cY_\m + y_\m , \t] ]
~\equiv ~\D \, \t ~.
\label{FP-operator}
\ee
Here, $\D$ is the Faddeev-Popov operator, denoted by $F[\F]$
in the previous section.
Let us introduce a generating functional $W[\f; J]$  by the rule
\be
{\rm e}^{{\rm i} W[\f;J]} =
 \int {\rm d} \vf \, {\rm Det} \D \,
{\rm e}^{ {\rm i} ( S[\f + \vf] + S_{\rm GF} [\c [\f; \vf ]]
+ J \cdot\vf)}~,
\label{gen-fun3}
\ee
with $J = (j_m, k_\m)$ the sources corresponding to
$ \vf =(a_m, y_\m)$, and define
\be
\langle \vf \rangle ~= ~\frac{\d}{\d J} \,  W[\f; J]~.
\ee
In terms of the Legendre transform of $W[\f; J]$,
\be
\G[\f; \langle \vf \rangle ]
= \big( W[\f; J] -  J \, \cdot \langle \vf \rangle
 \big)\big|_{J= J[\f; \langle \vf \rangle ]}~,
\ee
the effective action is
\be
\G[\f] =
\G[\f; \langle \vf \rangle =0] ~.
\ee

Using the conformal transformation laws (\ref{conf}),
the gauge condition $\c$ defined in (\ref{g-c}) changes as
 \cite{FP1,JKY}
\be
\d_{\rm c} \c = -\x \c - 2\s \c + 2(\pa^m \s) \,a_m
\equiv  \L \c + \hat{\d}_{\rm c} \c~, \qquad
\hat{\d}_{\rm c} \c = 2(\pa^m \s) \,a_m
\label{c-var}
\ee
under a combined conformal transformation
of the background and quantum fields, while
\be
\d_{\rm c}\, S_{\rm GF} [\c]
= -\frac{1}{ g^2} \int {\rm d}^4 x \;
{\rm tr} \, (\c \,\hat{\d}_{\rm c} \c) ~.
\ee
As can be seen,  the inhomogeneous part, $ \hat{\d}_{\rm c} \c$,
of the variation $\d_{\rm c} \c $
 makes $S_{\rm GF} [\c]$ conformally non-invariant.
Since $\pa_m \s = - 2b_m$, it is in fact the special conformal
transformations which render $S_{\rm GF} [\c]$ non-invariant.
${}$From (\ref{c-var}), one also observes that the Faddeev-Popov
determinant changes by the rule
\be
{\rm Det} \D  ~ \longrightarrow ~
{\rm Det} (\D + \hat{\d}_{\rm c} \D)~, \qquad
\hat{\d}_{\rm c} \D \, \t = -2(\pa^m \s) \,
(\cD_m \t +{\rm i}\, [a_m, \t] )~.
\ee
As a result, we have precisely the situation studied in the
previous section, and can therefore  make use
of the techniques described  there.

We will evaluate the variation
$W[\f +\d_{\rm c} \f; J] - W[\f;J]$ induced by a conformal
transformation $\f \to \f +\d_{\rm c} \f$ of the background fields,
with $\d_{\rm c} \f$ as in eq. (\ref{conf}), in the case when
$\langle \vf \rangle = 0$.
Using the path integral representation of $W[\f +\d_{\rm c} \f; J]$,
as in eq. (\ref{gen-fun3}), we change the integration variables
by the rule $\vf \to \vf + \d_{\rm c} \vf$. This gives
\be
{\rm e}^{{\rm i} W[\f +\d_{\rm c}\f;J]} =
 \int {\rm d} \vf \, {\rm Det} (\D + \hat{\d}_{\rm c} \D) \,
{\rm e}^{ {\rm i} ( S[\f + \vf]
+ S_{\rm GF} [\c + \hat{\d}_{\rm c} \c]
+ J \cdot\vf)}~.
\ee
The small deformation of the gauge conditions
in this expression can be compensated by a field dependent
gauge transformation, according to the rules described in
the previous section. This results in the
following conformal Ward identity
\bea
\d_{\rm c} \cA_m  \, \frac{\d \G[\f]} {\d \cA_m }
&+&\d_{\rm c} \cY_\m \,  \frac{\d \G[\f]} {\d \cY_\m } \non \\
=  - \langle D_m \, \frac{1} {\D} \,\hat{\d}_{\rm c} \c \rangle
\, \frac{\d   \G [\f; \langle \vf \rangle ]    }
{\d \langle a_m \rangle } \Big|_{ \langle \vf \rangle =0}
&+&{\rm i}\, \langle [ \frac{1} {\D} \,\hat{\d}_{\rm c} \c \, , \,
\cY_\m + y_\m ] \rangle \,
\frac{\d   \G [\f; \langle \vf \rangle ]    }
{\d \langle y_\m \rangle } \Big|_{ \langle \vf \rangle =0}~,
\label{conf-Ward}
\eea
with $D_m = \cD_m + {\rm i}\,a_m$. To complete the analysis,
we have to express
$ \d   \G [\f; \langle \vf \rangle ] / \d \langle \vf \rangle$
at $\langle \vf \rangle =0$ via $\d \G [\f] / \d \f$.

Given an infinitesimal change  $\d \f$
of the background fields, let us evaluate the variation
$W[\f +\d \f; J] - W[\f;J]$ at $\langle \vf \rangle =0$.
Using the path integral representation of $W[\f +\d \f; J]$,
as in eq. (\ref{gen-fun3}), we change the integration variables
in the manner $\vf \to \vf - \d \f$.
This gives
\be
{\rm e}^{{\rm i} W[\f +\d \f;J]} =
 \int {\rm d} \vf \, {\rm Det} (\D + \d \D) \,
{\rm e}^{ {\rm i} ( S[\f + \vf]
+ S_{\rm GF} [\c + \d \c]
+ J \cdot (\vf - \d \f))}~,
\label{aux}
\ee
where
\be
\d \c = - D^m \d \cA_m - {\rm i}\, [\cY_\m + y_\m ,
\d \cY_\m ]
\ee
and $\d \D$ is the deformation in $\D$ induced by $\d \c$.
The change of the gauge conditions in (\ref{aux}) can
be compensated by a field dependent gauge transformation,
as in the preceding section.
Then one gets (a similar relation was derived by Hart \cite{Hart})
\bea
&& \qquad \d \cA_m  \, \frac{\d \G[\f]} {\d \cA_m }
+\d \cY_\m \,  \frac{\d \G[\f]} {\d \cY_\m } \non \\
&=& \Big\{ \d \cA_m +
\langle D_m \, \frac{1} {\D} \,( D^n \d  \cA_n
+ {\rm i}\, [\cY_\n + y_\n , \d \cY_\n ] ) \rangle
\Big\} \, \frac{\d   \G [\f; \langle \vf \rangle ]    }
{\d \langle a_m \rangle } \Big|_{ \langle \vf \rangle =0}
\label{quantum-background}\\
&+& \Big\{ \d \cY_\m + {\rm i}\,
\langle [ \cY_\m  + y_\m \, , \,
\frac{1} {\D} \,( D^n \d  \cA_n
+ {\rm i}\, [\cY_\n + y_\n , \d \cY_\n ] ) \rangle \Big\}
\, \frac{\d   \G [\f; \langle \vf \rangle ]    }
{\d \langle y_\m \rangle } \Big|_{ \langle \vf \rangle =0}~.
\non
\eea
This relation allows
$ \d   \G [\f; \langle \vf \rangle ] / \d \langle \vf \rangle$
at $\langle \vf \rangle =0$ to be expressed in terms of $\d \G [\f] / \d \f$
and should be used in conjunction with the Ward identity
(\ref{conf-Ward}).
It should be pointed out that our conformal Ward identity
given by eqs. (\ref{conf-Ward}) and (\ref{quantum-background})
is more general than that derived in \cite{JKY}.

The conformal Ward identity can be obtained in the BRST approach
following the rules given in the previous section.
We do not pursue this approach here, but for completeness
give the expression for $S_{\rm eff}$ and the corresponding
BRST transformation. The action $S_{\rm eff}$ reads
\bea
S_{\rm eff} &=& = S[\f +\vf] + S_{\rm GF} [\c [\f; \vf ]]
+S_{\rm GH} [\f; \vf, {\bar C}, C]~, \non \\
S_{\rm GF} &=& -\frac{1}{ 2g^2} \int {\rm d}^4 x \;
{\rm tr} \, \c^2 ~, \qquad
S_{\rm GH} =  \int {\rm d}^4 x \;
{\rm tr} \,{\bar C} \, \D \, C~,
\eea
and is invariant under the {\it quantum} BRST transformation
\bea
\d a_m = -(\cD_m C +{\rm i}\, [ a_m , C]) \, \l ~, & \qquad &
\d y_\m = {\rm i} \, [C, \cY_\m +y_m]\, \l~, \non \\
\d C = -{\rm i}\, C^2 \, \l~, & \qquad &
\d {\bar C} = -\frac{1}{g^2}\, \c \, \l~.
\eea
The ghost fields are required to possess
the  conformal transformation laws \cite{FP1}
\be
-\d_{\rm c} {\bar C}  =  \x {\bar C}  + 2 \s {\bar C}~, \qquad
-\d_{\rm c} C =  \x C~,
\ee
and the same conclusion follows  from the analysis
of the preceding section.

\sect{One-loop calculations}

In this section, we compute  the leading quantum deformation
of the conformal transformation laws of the fields $\cA_m$ and
$\cY_\m$. Since there is an overall factor
of $1/g^2$ multiplying the action $S+ S_{\rm GF}$,
the loop-counting parameter is $g^2$.
We will examine the conformal Ward identity at
the one loop level or, equivalently, to order $g^2$.

${}$For the purpose of loop calculations,
 we  expand the action $S[\f +\vf]$
in powers of the quantum fields $\vf$ and
combine its quadratic part, $S_2$,
with the gauge fixing functional, $S_{\rm GF}$. This gives
\bea
S_2 + S_{\rm GF} = - \frac{1}{g^2} \int {\rm d}^4x \,{\rm tr}\,
\Big\{  \hf \, a^m \tilde {\D} a_m &+& {\rm i}\, \cF^{mn}[a_m, a_n]
+2 {\rm i}\,(\cD^m \cY_\m)\, [a_m, y_\m]
\non \\
 \qquad +  \hf \, y_\m \tilde{\D} y_\m
&-& [\cY_\m, \cY_\n] \, [y_\m , y_\n] \Big\}~,
\label{quadrat-action}
\eea
where $\cF_{mn}$ is the background field strength,
$[\cD_m, \cD_n] =  {\rm i}\, \cF_{mn}$, and
the operator $\tilde{\D}$,
\be
\tilde{\D} \,\t = - \cD^m\cD_m \,\t + [\cY_\m , [\cY_\m , \t ]] ~,
\ee
is simply the Faddeev-Popov operator $\D$ in eq.
(\ref{FP-operator}), evaluated at $a_m = y_\m =0$.
The last term in the right hand side of
(\ref{quadrat-action}) vanishes for an Abelian background.

In the action (\ref{quadrat-action}),
the trace is in the
fundamental representation
with the generators $T^{\rm F}_i$
normalized so that $ {\rm tr}\,(T^{\rm F}_i \,
T^{\rm F}_j) = \delta_{ij}.$
Later, we will have need to use the adjoint
representation $(T_i)_j{}^k = - \, {\rm i} \, f_{ij}{}^k$,
where the structure constants are defined by
$[T^{\rm F}_i, T^{\rm F}_j] = {\rm i} \,
f_{ij}{}^k \, T^{\rm F}_k.$
The adjoint representation matrices then satisfy
the normalization condition
${\rm tr}\,(T_i \, T_j) = 2N \, \delta_{ij}$
for gauge group $SU(N)$.

As mentioned earlier, the classical
$\cN=4$ super Yang-Mills action is derived by plain
dimensional reduction from the ten-dimensional super
Yang-Mills action (\ref{10Daction}).
To simplify  quantum calculations in the
$\cN=4$ super Yang-Mills theory,
it is convenient to restore ten-dimensional notation, as this
allows a unified treatment of $\cA_m$ and $\cY_\m$.
In ten-dimensional notation, the background fields are $\cA_M = (\cA_m,
\, \cY_\m)$ and the quantum
fields are $a_M = (a_m, \, y_\m).$
The full covariant derivatives $D_M$ are
defined by $D_M \phi = \partial_M \phi
+ {\rm i}\, [\cA_M + a_M, \phi]$,
with $\partial_M = (\partial_m, 0),$ and the background covariant
derivatives $\cD_M \phi = \partial_M \phi + {\rm i}\, [\cA_M, \phi]$
define the background field strength
$\cF_{MN}$ via $[\cD_M, \cD_N] = {\rm i}\, \cF_{MN}.$
The components of $\cF_{MN}$ are
$$ \cF_{mn} = \partial_m \cA_n - \partial_n \cA_m
+{\rm i}\,[\cA_m, \cA_n]~,
\qquad
\cF_{m \n} = \cD_m \cY_{\n}~, \qquad
\cF_{\m \n} = {\rm i}\,[\cY_{\m}, \cY_{\n}]~.$$
The gauge-fixing condition (\ref{g-c}) can be
expressed $\c = \cD^M a_M.$
The action (\ref{quadrat-action}) becomes
(with the ten-dimensional metric containing most pluses)
\bea
S_2 + S_{\rm GF}
& = & - \, \frac{1}{2g^2} \int {\rm d}^4x \,\,{\rm tr}\,
\Big\{   \, a^M \tilde{\D}\, a_M
- 2{\rm i}\, a_M \, [\cF^{MN}, \, a_N] \Big\}~,\non \\
&=&  - \, \frac{1}{2g^2} \int {\rm d}^4x
  \,\, a^{Mi} \, \Big( \tilde{\D} \,\delta_M{}^N
  - 2{\rm i}\, \cF_M{}^N
  \Big)_i{}^j \; a_{Nj}~,
\label{quadrat-action2}
\eea
where the operator $\tilde{\D}$ is just the covariant
d'Alembertian
\be
\tilde{\D} ~ = ~ - \cD^M\cD_M  ~.
\ee
In the second line of eq. (\ref {quadrat-action2}) and below,
we adopt a notation in which background fields are matrices in the adjoint
acting on quantum fields which are adjoint vectors.
${}$From eq. (\ref{quadrat-action2}),
we can read off the propagator
\be
\langle a_{Mi}(x) \, a^{Nj}(x') \rangle  = - {\rm i}\,
\Big( \frac{g^2}{\tilde {\D} - 2{\rm i}\, \cF }
\Big)_{Mi}{}^{Nj} \; \delta^4 (x,x')~.
\label{prop}
\ee

Now, we are prepared to analyse
quantum conformal invariance.
Using (\ref{c-var}), the conformal Ward identity (\ref{conf-Ward})
rewritten in ten-dimensional notation takes the form
\be
0 = \d_{\rm c} \cA_{Mi}  \, \frac{\d \G[\f]} {\d \cA_{Mi} }
+ 2\, (\partial^n \sigma) \, \langle
(D_M \, \D^{-1})_i{}^j \,
a_{nj} \rangle \, \frac{\d   \G [\f; \langle \vf \rangle ]    }
{\d \langle a_{Mi} \rangle } \Big|_{ \langle \vf \rangle =0}~.
\ee
As can be seen, the deformation of the conformal transformation law is
determined by the average
 $\langle (D_M \,  \D^{-1})_i{}^j \,
a_{nj} \rangle,$ which we will evaluate
at the one-loop level, i.e. to order $g^2$.
In this case, the quantum field $a_{nj}$ must be contracted
either
with a quantum field in $D_M$ or with a quantum field in
$\D^{-1}$, with the remaining (uncontracted) quantum fields
to be set to zero. The resulting conformal Ward identity is
\bea
0 ~=~
2\,{\rm i} \, (\partial^n \sigma) \,
(\delta_M{}^Q &+& \cD_M\,  \tilde{\D}^{-1} \, \cD^Q) \,
\langle a_Q^{k} \,
(T_k \, \tilde{\D}^{-1} )_i{}^j \,   a_{nj} \rangle \,
\frac{\d   \G [\f; \langle \vf \rangle ]    }
{\d \langle a_{Mi} \rangle } \Big|_{ \langle \vf \rangle =0} \non \\
&+& \d_{\rm c} \cA_{Mi}  \, \frac{\d \G[\f]} {\d \cA_{Mi} }
~+~ O(g^4)~,
\label{confSYM1}
\eea
with the group generators $T_k$ in the adjoint
representation.

The result in equation (\ref{quantum-background}) can now be used to
express the piece of (\ref{confSYM1}) containing
$ \d   \G [\f; \langle \vf \rangle ] / \d \langle a_{Mi} \rangle$
at $\langle \vf \rangle =0$ in terms of  $\d \G [\f] / \d \cA_{Mi}$.
Equation (\ref{quantum-background}) states
\be
( \delta_M{}^Q + \langle D_M \, \D^{-1} \, D^Q \,
\rangle ) \,\d  \cA_{Qi} \,
\frac{ \d   \G [\f; \langle \vf \rangle ]    }
{\d \langle a_{Mi} \rangle } \Big|_{ \langle \vf \rangle =0} =
\d \cA_{Mi}  \, \frac{\d \G[\f]} {\d \cA_{Mi} }~.
\label{conversion}
\ee
Since we are working only to order $g^2,$
if the order $g^2$ expression
$$
\d \cA_{Qi} = 2\,{\rm i} \, (\partial^n
\sigma) \,
\langle a_Q^{k} \,
(T_k \, \tilde{\D}^{-1} )_i{}^j \,   a_{nj} \rangle
$$
 is substituted into (\ref{conversion}), then the operator
$\bigl( \delta_M{}^Q + \langle D_M \, \D^{-1} \, D^Q \,
\rangle \bigr)$ acting on $\d \cA_Q$ is only required
at the tree level (order $g^0$),
in which case it becomes
$$
 \delta_M{}^Q +  \cD_M \, \tilde{\D}^{-1} \,
\cD^Q ~.
$$
This is precisely the
background-covariant transverse
{\it projection operator} which appears in the order
$g^2$ term in the conformal Ward identity
(\ref {confSYM1}). Thus we can use the result
(\ref{conversion}) to rewrite the conformal Ward identity
(\ref {confSYM1}) in the form
\be
0 ~=~ \Big\{ \d_{\rm c} \cA_{Mi}
+ 2\,{\rm i} \, (\partial^n \sigma) \,
\langle a_M^{k} \,
(T_k \, \tilde{\D}^{-1} )_i{}^j \,   a_{nj} \rangle \,\Big\}
\, \frac{\d \G[\f]} {\d \cA_{Mi} } ~+~ O(g^4)~.
\label{finalident}
\ee

It remains to compute
$$\langle a_M^{k} \,
(T_k \, \tilde{\D}^{-1} )_i{}^j \,   a_{nj} \rangle ~.
$$
Since in the adjoint representation $(T_k)_i{}^j =
-{\rm i} \, f_{ki}{}^j = - (T_i)_k{}^j$, this can be expressed as
$$
 - (T_i \,\tilde{\D}^{-1} )_k{}^j \,  \langle a_{nj} \, a_M^k
\rangle ~,
$$
where there is an implicit functional trace.
Expanding the propagator (\ref{prop}) in powers of the background
field strength,
\be
\langle a_M^{k} \,
(T_k \, \tilde{\D}^{-1} )_i{}^j \,   a_{nj} \rangle
= {\rm i} g^2\, \eta_{Mn} \, {\rm tr} \,
( T_i \tilde{\D}^{-2}) - 2 g^2\,
\,
{\rm tr}\, (T_i \tilde{\D}^{-2} \, \cF_{nM} \,
\tilde{\D}^{-1}) + O(\cF^2)~.
\label{vev}
\ee

Up to this point, the background fields have been completely arbitrary.
${}$From now on, we take the gauge group to be $SU(N+1)$
and restrict  attention to a specific Abelian
background which is of interest in the context of the AdS/CFT
duality. For a single D3-brane probe separated from a stack of $N$ D3-branes,
the relevant background is
\be
\cA_m = A_m \, T_0~, \qquad
\cY_\m = Y_\m \, T_0~,
\ee
where $A_m$ is the $U(1)$ gauge field on the world-volume of
the probe brane, and the world-volume scalars $Y_\m$
are the components of the transverse
separation of the probe brane from the stack of branes.
The adjoint generator $T_0$ corresponds
to the following $ su(N+1) $  generator
$$
T^{\rm F}_0 =  \frac{1}{\sqrt{N(N+1)}} \,
{\rm diag}\, ( -1,  \cdots, -1, N)~,
$$
and its explicit form is
$$
T_0 =  \sqrt{\frac{N+1}{N}}\, {\rm diag} \,
( 0, \cdots, 0, 1, -1, \cdots, 1, -1)~,
$$
where there are $N^2$ zeroes on the diagonal and $N$ pairs $(1,-1).$

${}$For the background chosen,
the first term on the right hand side of (\ref{vev}) can be
neglected;
as will be shown later, it leads to a pure gauge transformation.
Let us therefore concentrate on the second term.
Restricting attention to modifications to the conformal transformation
of $\cA_M$ which contain at most one derivative
(and hence the $O(\cF^2)$ corrections in (\ref{vev}) can be neglected),
 the functional trace is expressed in momentum space as
$$ \int \frac{{\rm d}^4 k}{(2 \pi)^4} \, {\rm tr} \,
\Big( T_i \,
\frac{1}{(k^2 + \cY_{\n}\cY_\n)^2} \,\, \cF_{nM} \,\,
\frac{1}{(k^2 + \cY_{\n}\cY_\n)} \Big)~.
$$
The result in this case is that, using (\ref{finalident}), the
leading  quantum deformation of the conformal
transformation properties of $\cA_{Mi}$ is
$$
 \hat{\delta}_{\rm c} \cA_{Mi}
 =   \frac{g^2}{8 \pi^2} \, (\partial^n \sigma)
\frac{F_{nM}}{Y^2} \, \frac{N}{N+1} \, {\rm
tr}\, (T_i T_0)~.
$$
The trace is in the adjoint representation, and ensures that only the
nonvanishing components of the background receive a quantum
modification:
\be
 \hat{\delta}_{\rm c} A_M =  \frac{N g^2}{4 \pi^2} \, (\partial^n \sigma)
\frac{F_{nM}}{Y^2}~.
\ee
Reducing to four-dimensional notation, this yields
\be
\hat{\delta}_{\rm c} A_m
= - \, \frac{N g^2}{4 \pi^2} \, (\partial^n \sigma)
\frac{F_{mn}}{Y^2}~, \qquad
\hat{\delta}_{\rm c} Y_\m =  \frac{N g^2}{4 \pi^2} \, (\partial^n \sigma)
\frac{\partial_n Y_\m}{Y^2}~.
\label{def-transf}
\ee
The deformation $\hat{\delta}_{\rm c} Y_\m$ was  computed previously
in \cite{JKY}.

Let us finally
return to the first term on the right hand side of (\ref{vev}).
It results in a
deformation to  the conformal transformation of
$\cA_{mi}$  of the form
$-\, 2 \, g^2 \, (\partial_m \sigma) \, {\rm tr}\,  ( T_i
\tilde{\D}^{-2})$.
If  only the terms which are at most linear
in derivatives of the background fields are retained,
this  can be expressed in momentum space as
\bea
(\partial_m \sigma) \,
\int \frac{{\rm d}^4 k}{(2\pi)^4} \,
{\rm tr}\, \Big( T_i \,
\frac{1}{(k^2 + \cY_\n \cY_\n)^2}\Big ) &=&
\pa_m \Big\{ \sigma
\int \frac{ {\rm d}^4 k}{(2\pi)^4} \, {\rm tr} \,
\Big( T_i \,
\frac{1}{(k^2 + \cY_\n \cY_\n)^2} \Big) \Big\}
\non \\
&+ &4
\sigma
\int \frac{{\rm d}^4k}{(2\pi)^4} \, {\rm tr}
\Big(T_i \, \cY_\n \pa_m
\cY_\n \,  \frac{1}{(k^2 + \cY_\n \cY_\n)^3}\Big)~.\non
\eea
The momentum integral in the first term on
the right hand side is ultraviolet
divergent; however, the overall derivative means that the
contribution is pure gauge and so can be ignored.
The second term vanishes on group theoretic grounds,
because the adjoint represenation is
non-chiral, ${\rm tr} \, (T_i \,\{T_j , T_k\} ) = 0$.

\sect{Discussion}

In order to put the result (\ref{def-transf}) in the
context of AdS/CFT correspondence,
consider the bosonic action of a
single D3-brane probe moving near the core of the stack of
$N$  D3-branes
(we set $2\p \a' =1$  and ignore the Chern-Simons term,
see, e.g. \cite{Ts} for more detail):
\be
S = - T_3 \int {\rm d}^4x
\left( \sqrt{ - {\rm det} \Big( \frac{Y^2}{R^2} \, \eta_{m
n} +  \frac{R^2}{Y^2}\, \partial_{m}Y_\m \partial_{n} Y_\m
+ F_{mn} \Big)} - \frac{Y^4}{R^4}~
\right) ~,
\label{d3brane}
\ee
where $Y^2 = Y_\m Y_\m$,
$T_3 = 1/ g^2$
and  $R^4 = N\, g^2 / (2 \p^2)$.
The action is invariant under the $AdS_5 \times S^5$
field transformations \cite{M,Kallosh}
\bea
\d A_m &=& \d_{\rm c} A_m
-\frac{R^4}{2Y^2}\, (\pa^n \s) \, F_{mn}
+ \pa_m \Big( \frac{R^4}{2 Y^2}\, (\pa^n \s)\, A_n \Big)  ~,
\label{A-def} \\
\d Y_\m &= &\d_{\rm c} Y_\m
+ \frac{R^4}{2 Y^2}\, (\pa^n \s) \, \pa_n Y_\m~,
\label{Y-def}
\eea
with $\d_{\rm c} A_m $ and $ \d_{\rm c} Y_\m$
the linear conformal transformations (\ref{conf}).
The nonlinear terms in (\ref{A-def}) and
(\ref{Y-def}) coincide with the quantum deformation
(\ref{def-transf}) except for the total derivative
in (\ref{A-def}). Of course, the latter term is not essential
since it generates a pure gauge transformation.
However, only if it is retained do
the variations $\d A_m $ and $\d Y_\m$ provide
a representation of the conformal algebra.

Eq. (\ref{def-transf}) constitutes the leading quantum
deformation, in the framework of the loop expansion
and the derivative expansion, to the conformal
transformations of $A$ and $Y$. It would be interesting
to analyse higher loop and higher derivative deformations.
Of course, the fermions and the ghosts have to be taken into
account at higher loops.
Independently of what happens
at higher loops, the one-loop deformation (\ref{def-transf})
is a remarkable result. In conjunction with the requirement
of $SO(6)$ $R$-symmetry
and some non-renormalization theorems in $\cN=4$ SYM,
the conformal transformation
(\ref{Y-def}) is known to uniquely fix the action
(\ref{d3brane}) for $F_{mn}= 0$ \cite{M}.
On the super Yang-Mills side, this action  results from
summing up quantum corrections to all loop orders.
We believe that the deformed conformal invariance
in the $\cN=4$ super Yang-Mills theory should be
crucial, along with the requirement
of nonlinear self-duality \cite{GKPR,KT},
for a better understanding  of numerous non-renormalization
theorems which are predicted by the AdS/CFT conjecture
and relate to the explicit structure of the
low energy effective action in $\cN=4$ super Yang-Mills
theory (see \cite{CT,BKT,BPT} for a more detailed discussion
and  additional references).

Along with  quantum loop calculations,
there is a purely field theoretic  problem
to classify possible {\it local} field-dependent
deformations of the classical conformal transformation,
given by $\d_{\rm c} A$ and $\d_{\rm c} Y$, such that
they provide a nonlinear realization of the conformal
algebra. This is an interesting and challenging problem
which may be addressed within  the local BRST cohomological
approach (see \cite{BBH} for a review).
It seems that the AdS deformation defined by eqs.
(\ref{A-def}) and (\ref{Y-def}) is the only
nontrivial solution to first order in derivatives.
Of course, a similar problem may be formulated in terms
of superfields.
In the case of $\cN =1$ supersymmetry, for example,
one can start with an Abelian gauge superfield  $V$ and
three chiral superfields $\F^i$ and then try to deform
their linear superconformal transformations
(chosen to leave the free actions invariant)
so as to end up with an
analogue of eqs. (\ref{A-def}) and (\ref{Y-def}).
${}$For a single chiral scalar superfield $\F$,
this problem has been  solved implicitly
in \cite{KM3}.

The $\cN=4$ super Yang-Mills theory can be formulated in
$\cN=1$ or $\cN=2$ superspaces.
It is therefore natural to wonder
whether there may be efficient superfield extensions of
the approach advocated in the present paper.
The $\cN=1$ superfield formulation,
although most familiar, does not seem to be useful.
The point is that the Yang-Mills gauge transformations
are known to be highly nonlinear in $\cN=1$ superspace.
In addition, there is no simple $\cN=1$ superfield
generalization\footnote{The supersymmetric $R_\x$ gauge,
which was introduced in \cite{OW} and further studied in \cite{BSS},
is nonlocal and, therefore, a special care
is required to make (an extansion of) this gauge
useful for practical calculations within the $\cN=1$
background field scheme.}
of 't Hooft's gauge, without  which
one inevitably runs into ugly infrared problems.
More promising is the $\cN=2$
harmonic superspace formulation (see \cite{GIOS} for a review),
which is similar, in several respects,
to the ordinary component formulation.
In $\cN=2$ harmonic superspace,
one has a well elaborated background field method
\cite{N=2BFM} and quite powerful heat kernel techniques
\cite{KM}. The price to pay here, however, might be
the need to be extremely careful when evaluating
the relevant supergraphs in order to avoid the appearance
of so-called harmonic singularities.

\vskip.5cm

\noindent
{\bf Acknowledgements.}
Discussions with Arkady Tseytlin are gratefully acknowledged.
We are grateful to Joseph Buchbinder for bringing references
\cite{OW,BSS} to our attention.
This work is partially supported by a University of Western
Australia Small Grant and an ARC Discovery Grant.


\begin{thebibliography}{99}

\bibitem{M}
J.~Maldacena, Adv. Theor. Math. Phys. {\bf 2} (1998) 231
[hep-th/9711200].

\bibitem{GKP}
S.S.~Gubser, I.R.~Klebanov and A.M.~Polyakov,
Phys.\ Lett.\ {\bf B428} (1998) 105 [hep-th/9802109].

\bibitem{Witten}
E.~Witten,
Adv.\ Theor.\ Math.\ Phys.\  {\bf 2} (1998) 253
[hep-th/9802150].

\bibitem{AGMOO}
O.~Aharony, S.S.~Gubser, J.~Maldacena, H.~Ooguri and Y.~Oz,
Phys.\ Rept.\  {\bf 323} (2000) 183
[hep-th/9905111].

\bibitem{JKY}
A.~Jevicki, Y.~Kazama and T.~Yoneya,
Phys.\ Rev.\ Lett.\  {\bf 81} (1998) 5072
[hep-th/9808039];
Phys.\ Rev.\ {\bf D59} (1999) 066001
[hep-th/9810146].

\bibitem{FP1}
E.S.~Fradkin and M.Y.~Palchik,
Phys.\ Lett.\ {\bf B147} (1984) 86.

\bi{Pal} M.Ya. Palchik,
in {\it Quantum Field Theory and Quantum Statistics},
I.A. Batalin, C.J. Isham and G.A. Vilkovisky (Eds.),
Adam Hilger, Bristol, 1987, Vol.1, p. 313.

\bibitem{FP2}
E.S.~Fradkin and M.Y.~Palchik,
{\it Conformal Quantum Field Theory in D-Dimensions},
Kluwer, Dordrecht, 1996;
Phys.\ Rept.\  {\bf 300} (1998) 1.

\bibitem{BV}
I.A.~Batalin and G.A.~Vilkovisky,
Phys.\ Lett.\ {\bf B102} (1981) 27;
Phys.\ Lett.\ {\bf B120} (1983) 166;
Phys.\ Rev.\ {\bf D28} (1983) 2567
[Erratum-ibid.\ {\bf D30} (1983) 508].

\bibitem{BHW}
F.~Brandt, M.~Henneaux and A.~Wilch,
Phys.\ Lett.\ {\bf B387} (1996) 320 [hep-th/9606172].

\bibitem{vH}
J.W.~van Holten,
Phys.\ Lett.\ {\bf B200} (1988) 507.

\bibitem{dWF}
B.~de Wit and D.Z.~Freedman,
Phys.\ Rev.\ {\bf D12} (1975) 2286.


\bi{DeWitt67}
B.C.~DeWitt,
Phys. Rev. {\bf 162} (1967) 1195.

\bibitem{DeWitt2}
B.S.~DeWitt,
in {\it Les Houches 1983: Relativity, Groups and Topology II},
B.S. DeWitt and R. Stora (Eds.), Elsevier, Amsterdam, 1984,
p. 381.

\bibitem{DeWitt1}
B.S.~DeWitt,
in {\it Recent Developments in Gravitation,
Cargese 1978}, M. Levy and S. Deser (Eds.),
Plenum Press, New York, 1978,  p. 275.

\bi{NK}
N.K.~Nielsen,
Nucl.\ Phys.\ {\bf B140} (1978) 499; \\
R.E. Kallosh, Nucl.\ Phys.\ {\bf B141} (1978) 141.

\bi{Wei} S. Weinberg, {\it The Quantum Theory of Fields},
Vol. II, Cambridge University Press, Cambridge, 1996.

\bibitem{BRST}
C.~Becchi, A.~Rouet and R.~Stora,
Annals Phys.\  {\bf 98} (1976) 287;\\
I.V. Tyutin, Lebedev FIAN preprint 39 (1975).

\bibitem{BFV}
E.S.~Fradkin and G.A.~Vilkovisky,
Phys.\ Lett.\ {\bf B55} (1975) 224;
Lett.\ Nuovo Cim.\  {\bf 13} (1975) 187;
I.A.~Batalin and G.A.~Vilkovisky,
Phys.\ Lett.\  {\bf B69} (1977) 309; \\
E.S.~Fradkin and T.E.~Fradkina,
Phys.\ Lett.\ {\bf B72} (1978) 343.

\bibitem{tH} G. 't Hooft,
in Proceedings of the {\it 12th Winter School of Theoretical Physics in
Karpacz}, Acta Univ. Wratislaviensis {\bf 368} (1976) 345.

\bibitem{DeWitt3} B.S. DeWitt,
in {\it Quantum Gravity II},  C. Isham, R. Penrose
and D. Sciama (Eds.), Oxford University Press, New York, 1981,
p. 449.

\bibitem{Ab} L.F. Abbott,
Nucl. Phys. {\bf B185} (1981) 189;
Acta Phys. Polon. {\bf B13} (1982) 33.

\bi{Boul} D.G. Boulware, Phys. Rev. {\bf D23} (1981) 389.

\bi{Hart} C.F. Hart, Phys. Rev. {\bf D28} (1983) 1993.

\bi{Ts} A.A.~Tseytlin, {\it Born-Infeld action, supersymmetry and
string theory}, hep-th/9908105.

\bibitem{Kallosh}
R.~Kallosh, J.~Kumar and A.~Rajaraman,
Phys.\ Rev.\ {\bf D57} (1998) 6452
[hep-th/9712073];
P.~Claus, R.~Kallosh, J.~Kumar, P.K.~Townsend and A.~Van Proeyen,
JHEP {\bf 9806} (1998) 004
[hep-th/9801206].

\bi{GKPR} F. Gonzalez-Rey, B. Kulik, I.Y. Park and
M. Ro\v{c}ek, Nucl. Phys. {\bf B544} (1999) 218
[hep-th/9810152].

\bi{KT}
S.M.~Kuzenko and S.~Theisen,
JHEP {\bf 0003} (2000) 034
[hep-th/0001068].

\bibitem{CT}
I.~Chepelev and A.A.~Tseytlin,
Nucl.\ Phys.\  {\bf B511} (1998) 629[hep-th/9705120];
Nucl.\ Phys.\ {\bf B515} (1998) 73 [hep-th/9709087].

\bibitem{BKT}
I.L.~Buchbinder, S.M.~Kuzenko and A.A.~Tseytlin,
Phys.\ Rev.\  {\bf D62} (2000) 045001 [hep-th/9911221].

\bibitem{BPT}
I.L.~Buchbinder, A.Y.~Petrov and A.A.~Tseytlin,
Nucl.\ Phys.\  {\bf B621} (2002) 179 [hep-th/0110173].

\bibitem{BBH}
G.~Barnich, F.~Brandt and M.~Henneaux,
Phys.\ Rept.\  {\bf 338} (2000) 439
[hep-th/0002245].

\bibitem{KM3}
S.M.~Kuzenko and I.N.~McArthur,
Phys.\ Lett.\  {\bf B522} (2001) 320 [hep-th/0109183].

\bibitem{OW}
B.A.~Ovrut and J.~Wess,
Phys.\ Rev.\ {\bf D25} (1982) 409.

\bibitem{BSS}
P.~Bin\'etruy, P.~Sorba and R.~Stora,
Phys.\ Lett.\  {\bf B129} (1983) 85.

\bibitem{GIOS}
A.S.~Galperin, E.A.~Ivanov, V.I.~Ogievetsky and E.S.~Sokatchev,
{\it Harmonic Superspace},
Cambridge University Press, Cambridge, 2001.

\bibitem{N=2BFM}
I.L.~Buchbinder, E.I.~Buchbinder, S.M.~Kuzenko and B.A.~Ovrut,
Phys.\ Lett.\ {\bf B417} (1998) 61 [hep-th/9704214];
I.L.~Buchbinder, S.M.~Kuzenko and B.A.~Ovrut,
Phys.\ Lett.\ {\bf B433} (1998) 335 [hep-th/9710142];
I.L.~Buchbinder and S.M.~Kuzenko,
Mod.\ Phys.\ Lett.\  {\bf A13} (1998) 1623 [hep-th/9804168].

\bibitem{KM}
S.M.~Kuzenko and I.N.~McArthur,
Phys.\ Lett.\  {\bf B506} (2001) 140 [hep-th/0101127];
Phys.\ Lett.\ {\bf B513} (2001) 213 [hep-th/0105121].


\end{thebibliography}
\end{document}